\documentstyle[12pt]{article}

\begin{document}
\newcommand{\dia}{\begin{displaymath}}
\newcommand{\die}{\end{displaymath}}
\newcommand{\eqa}{\begin{equation}}
\newcommand{\eqe}{\end{equation}}
\newcommand{\eqna}{\begin{eqnarray}}
\newcommand{\eqne}{\end{eqnarray}}
\newcommand{\eqnaa}{\begin{eqnarray*}}
\newcommand{\eqnae}{\end{eqnarray*}}
\newcommand{\fraz}{\frac{1}{2}}
\newcommand{\frav}{\frac{1}{4}}
\newcommand{\frasz}{\frac{1}{\sqrt{2}}}
\newcommand{\tr}[1]{\mbox{Tr}\left\{#1\right\}}
\newcommand{\labe}[1]{\label{#1}}

\begin{titlepage}
\title{Gauge fixing by unitary transformations in QCD}
\author{Dieter Stoll\\ Department of Physics, Faculty of Science,
University of Tokyo\\ Bunkyo--ku, Tokyo 113, Japan\thanks{email:
 stoll@tkyux.phys.s.u-tokyo.ac.jp\hspace{.5cm}FAX: (81)(3)56849642}}
\date{  }
\maketitle
\begin{abstract}
The unitary gauge fixing technique is applied to the QCD hamiltonian
formulated in terms of angular variables. It is demonstrated that
in this formulation projections on the physical Hilbert space are
unnecessary to separate physical and unphysical degrees of freedom.
Therefore the application of the unitary gauge fixing technique
can be extended to the operator level.
\end{abstract}

\end{titlepage}

The development of an analytical understanding of non--perturbative aspects
of QCD is still one of the most challenging tasks. As a step towards this
goal the interest in formulations of QCD, in which the gauge
invariant variables are explicitly isolated, has been renewed recently.
After the pioneering work of Goldstone and Jackiw \cite{Goldstone} different
such formulations have been derived \cite{Baluni}--\cite{Lenz} and on this
basis an attempt to calculate approximately the spectrum of low lying states
has been made \cite{Martin}. Most of these aforementioned approaches are
based on a particular ad hoc separation of gauge field degrees of freedom
into gauge variant angular variables and gauge invariant variables. More
general methods of eliminating unphysical variables, however, have also
been developed \cite{Christ,Lenz,Ohta}. They are particularly
successful in QED, where unitary transformations within the physical
subspace can be obtained, the property of which is to relate formulations in
different gauges without using unphysical, gauge variant degrees of freedom
\cite{Ohta}. In the application of the unitary gauge fixing technique
to QCD, however, this generality could not be maintained. Instead explicit
projections onto the physical subspace were necessary in order to derive
one particular representation of the QCD hamiltonian, the axial gauge
representation \cite{Lenz}. This presents an obstacle not only for establishing
the aforementioned relation between different gauges, but also obscures the
important issue of residual symmetries which has been linked to the presence
of massless exitations \cite{Lenz,Thies}. In this letter we want to demonstrate
that these limitations do not arise, if the hamiltonian is written in terms
of angular variables. Introducing angular degrees of freedom explicitely,
allows us to perform the unitary gauge fixing transformations on the operator
level. To demonstrate this possibility we reproduce the axial gauge
representation of the QCD hamiltonian which was formulated in \cite{Lenz}.

We use the hamiltonian formulation in which the dynamics is defined in terms
of the standard canonically quantized hamiltonian in the $A_0=0$ gauge
\eqa {\cal H}=\sum_i\bar\psi(x)\gamma_i\left(i\partial_i+gA_i\right)\psi(x)
+m\bar\psi(x)\psi(x) +\fraz \sum_i E_i^a(x)E_i^a(x)+ \frac{1}{4} \sum_{ij}
F_{ij}^aF_{ij}^a \labe{hnot}\eqe
supplemented by the contraint that the generators of time independent gauge
transformations, Gauss' law operators $G^a$, should annihilate physical states
\eqa G^a(x)=\sum_i\left[\partial_iE_i^a(x)+gf^{abc}A_i^b(x)E_i^c(x)\right]
+g\rho(x);\quad G^a(x)|phys>=0\  .\eqe
In these expressions space is assumed to be a torus of volume $L^d$ and the
gauge fields are canonically quantized with periodic, fermion fields with
anti-periodic boundary conditions. The gauge field may be written in SU(N)
gauge theories as\footnote{Note that we deviate from the usual convention
in that spatial indices are summed over only, if explicitely indicated.}
\eqa A_i(x)=\frac{i}{g}V_i(x)\partial_iV_i^{\dag}(x) \quad (\mbox{no
summation})
\labe{afeld}\eqe
where the SU(N)-matrices $V_i$ are uniquely determined, given a definite set
of paths which we choose to be straight lines parallel to one of the coordinate
axes
\eqa V_i(x)=P\exp\left[ig\int_0^{x_i}dz_iA_i(x_i^\perp,z_i)
\right]=\exp\left[i\xi_i^a(x)\frac{\lambda^a}{2}\right]\labe{xidef}\ .\eqe
P in this expression denotes path ordering and $x_i^\perp$ stands for all
coordinates orthogonal to $x_i$. The matrices $V_i(x)$ are furthermore
parametrized in terms of "angles" $\xi_i(x)$. In order to formulate the
hamiltonian using this new set of variables we also introduce angular
momentum operators $J_k^c(x)$ with commutation relations
\eqna \left[J_i^a(x),V_j(y)\right]&=&\delta_{i,j}\delta^d(x-y)V_j(y)
\frac{\lambda^a}{2}\labe{jvcom}\\   \left[J_i^a(x),J_j^b(y)
\right]&=&\delta_{i,j}if^{abc}J_i^c(x)\delta^d(x-y)\nonumber \eqne
in terms of which the electric field operator can be written as
\eqnaa E_i^a(x) &=& g\int d^dz \delta^{d-1}(z_i^\perp-x_i^\perp)\theta(z_i-x_i)
\theta(x_i)N_i^{ac}(x) J_i^c(z) \\  N_i^{ac}(x) &=& \tr{V_i(x)
\frac{\lambda^a}{2}V_i^{\dag}(x)\lambda^c}\ . \eqnae
Using these relations one easily rewrites the hamiltonian in the form (an
analogous expression was derived in \cite{Bars})
\eqna {\cal H}&=&\sum_i\left[\bar\psi(x)V_i(x)\right]\gamma_ii\partial_i\left[
V_i^{\dag}(x)\psi(x)\right]+m\bar\psi(x)\psi(x)\nonumber \\ & +& \frac{g^2}{2}
\sum_i\int_{x_i}^Ldz_i J_i^b(x_i^\perp,z_i)\int_{x_i}^Ldz_i^\prime
J_i^b(x_i^\perp,z_i^\prime) \labe{hneu}\\
&+& \frac{1}{2g^2}\sum_{ij} \tr{\left[\partial_i\left[\left(
V_i^{\dag}V_j\right)\partial_j\left(V_j^{\dag}V_i\right)\right]\right]\left[
\partial_i\left[\left(V_i^{\dag}V_j\right)
\partial_j\left(V_j^{\dag}V_i\right)\right]\right]^{\dag}}\nonumber \eqne
and for the Gauss' law operators we obtain the following expressions
\eqa G^a(x)=-g\sum_i \left[ N_i^{ac}(x) J_i^c(x)- \delta(x_i)\int dz_i
J_i^a(x_i^\perp,z_i)\right] + g\rho^a(x) \label{gauss}\ .\eqe
The technique to use unitary transformations in order to eliminate unphysical
degrees of freedom \cite{Lenz,Ohta} is, in the case of the axial gauge
$A_3^a=0$, based on the use of the transverse Gauss' law operator $G_\perp$ in
terms of which the following unitary transformation $T$ was
constructed in \cite{Lenz}
\eqna G_\perp^a(x) &=& -g\sum_{i\neq 3}\left[ N_i^{ac}(x) J_i^c(x)-
\delta(x_i)\int dz_i J_i^a(x_i^\perp,z_i)\right] + g\rho^a(x) \label{gauss1}\\
T &=& \exp\left[\frac{i}{g}\int
d^dzG_\perp^{a_0}(z) \theta^{a_0}(z_3^\perp)\frac{z_3}{L}\right]\exp\left[
\frac{-i}{g}\int d^dzG_\perp^a(z)\Delta^a(z_3^\perp)\right]\\  &&
\exp\left[ \frac{-i}{g}\int d^dzG_\perp^a(z)\xi_3^a(z)\right] \ .\eqne
$\Delta(x_3^\perp)$ and $\theta(x_3^\perp)$ in these expressions are defined
by diagonalizing $\xi_3(x_3^\perp,L)$\footnote{Note that
we adopt the convention that indices with subindex 0
enumerate the elements of the Cartan subalgebra of SU(N) and indices
with subindex 1  enumerate all remaining elements. }
\eqna \exp\left[-i\Delta(x_3^\perp)\right]\exp\left[i\xi_3(x_3^\perp,L)\right]
\exp\left[i\Delta(x_3^\perp)\right] &=& \exp\left[i\theta(x_3^\perp,L)
\right]\labe{v3def}\\  \exp\left[-i\Delta(x_3^\perp)\right]\frac{\lambda^a}{2}
\exp\left[i\Delta(x_3^\perp)\right] &=& \sum_b R_{ab}(x_3^\perp)
\frac{\lambda^b}{2} \labe{rdef}\\ \tr{e^{-i\theta(x_3^\perp)}
\frac{\lambda^{a_1}}{2} e^{i\theta(x_3^\perp)}\lambda^{b_1}}-\delta_{a_1b_1}
&=&  P_{a_1b_1}(x_3^\perp)\labe{pdef}\ .\eqne
For later use we have already introduced the matrices $R(x_3^\perp)$ and
$P(x_3^\perp)$.  Taking notice of the following general relation
which is valid for any function $\varphi$ of the "angles" $\xi_3(x)$
\dia \exp\left[\frac{-i}{g}\int d^dzG_\perp^a(z)\varphi^a(z)\right]
\frac{-i\delta}{\delta\xi_3^b(x)}
\exp\left[\frac{i}{g}\int d^dzG_\perp^a(z)\varphi^a(z)\right]=\die
\eqa \frac{-i\delta}{\delta\xi_3^b(x)} +\frac{2}{g}\int d^dz\tr{G_\perp(z)
e^{-i\varphi(z)}\frac{-i\delta}{\delta\xi_3^b(x)}e^{i\varphi(z)}}\eqe
we can unitarily transform $J_3^a(x)$ and $G^a(x)$\footnote{Exept these two
all variables associated with \protect$A_{i\neq 3}$ are, by construction, just
gauge transformed with the transformation function depending on
\protect$\xi_3(x)$.}. We find the following result
\eqna TgJ_3^a(x)T^{\dag} &=& gJ_3^a(x)+R_{ac_0}(x_3^\perp)\left[
G_\perp^{c_0}(x)-\delta(x_3-L)\int dz\frac{z}{L}G_\perp^{c_0}
(x_3^\perp,z) \right]\\ &&+ R_{ad_1}(x_3^\perp)\left[{\cal G}^{d_1}(x)
+\delta(x_3-L)P_{d_1c_1}^{-1}(x_3^\perp)\int dz
{\cal G}^{c_1}(x_3^\perp,z)\right] \nonumber \\  TG^a(x)T^{\dag} &=&
-g N_3^{ac}(x) J_3^c(x)+ \delta(x_3)\int dz \left[gJ_3^a(x_3^\perp,z)
+R_{ac_0}(x_3^\perp)G_\perp^{c_0}(x_3^\perp,z)\right]\nonumber\ . \eqne
Upon decomposing $J_3^a(x)$ with the use of a $\delta$--function in the
following way
\eqa J_3^a(x)= \frac{1}{L}\sum_n J_3^a(x_3^\perp,n) \left[e^{i2\pi nx_3/L}-
1\right] +\frac{1}{L}\sum_ne^{i2\pi nx_3/L}J_3^a(x_3^\perp,L)\labe{jres}\ .\eqe
one can separate the Gauss' law constraints into three independent
contributions
\eqna 0<x_3<L\ :\quad  J_3^a(x)|phys>&=& 0 \\
R_{ab_1}(x_3^\perp)J_3^a(x_3^\perp,L)|phys.> &=& 0 \labe{jcon} \\
\int_0^L dz G_\perp^{b_0}(x_3^\perp,z)|phys.> &=& 0 \labe{tcon}\eqne
the last of which eq.(\ref{tcon}) generates residual abelian gauge
transformations. These can be implemented without difficulty \cite{Lenz} and
for this reason they shall not be considered here any further.

Having identified the independent constraints, the hamiltonian
can be decomposed into a part which is proportional
to Gauss' law operators and thus vanishes on physical states and
a remaining hamiltonian acting non--trivially in the physical Hilbert
space. For the latter part we find using the identities
given in the appendix
\eqa T\left[\int dx_3\int_{x_3}^L dy \int_{x_3}^L dz \frac{g^2}{2}
J_3^a(x_3^\perp,y)J_3^a(x_3^\perp,z)\right] T^{\dag} \stackrel{phys.space}{=}
\frac{g^2L}{2} J_3^a(x_3^\perp,L) J_3^a(x_3^\perp,L)  \eqe
\eqna &+& \fraz
 \int dy\int dz \left[\theta(y-z)z+ \theta(z-y)y\right]
{\cal G}^{c_1}(x_3^\perp,y){\cal G}^{c_1}(x_3^\perp,z)\nonumber \\ &+& \frac{
L}{2} P_{c_1a_1}^{-1}(x_3^\perp)\int dy\int dz\left[\left(P_{c_1b_1}^{-1}
(x_3^\perp) +\frac{y}{L}\delta_{c_1,b_1} \right) \delta_{a_1,d_1}
+\frac{z}{L}\delta_{c_1,d_1}\delta_{a_1,b_1}\right]{\cal G}^{b_1}(x_3^\perp,y)
{\cal G}^{d_1}(x_3^\perp,z) \nonumber \\  &-&  \fraz \int dy\int dz\left[
\frac{yz}{L}-\theta(y-z)z-\theta(z-y)y\right]G_\perp^{c_0}(x_3^\perp,y)
G_\perp^{c_0}(x_3^\perp,z) \eqne
where we note that the contribution $J_3^a(x_3^\perp,L)
J_3^a(x_3^\perp,L)$ has not yet been reduced to the relevant part
in the physical Hilbert space since eq.(\ref{jcon}) has not yet been used.
The contribution in the physical subspace depending on $G_\perp$
can be brought into the following form
\eqna H_{int} &=& \int d^{d-1}x_3^\perp\int dy\int
dz\left[-\frac{|y-z|}{4}+ \frac{(y-z)^2}{4L}\right]
G_\perp^{c_0}(x_3^\perp,y)G_\perp^{c_0}(x_3^\perp,z)\nonumber \\ &+&
\frac{L}{2} \int d^{d-1}x_3^\perp \int dy\int dz \sum_{nm}{\cal G}_{nm}
(x_3^\perp,y) {\cal G}_{mn}(x_3^\perp,z)\left[\sin^{-2}\left[\frac{
\mu_n(x_3^\perp) -\mu_m(x_3^\perp)}{2}\right]\right. \nonumber \\
&&\left.+ i\frac{2}{L} \cot\left[\frac{\mu_n(x_3^\perp)-\mu_m(x_3^\perp)}{2}
\right](y-z) -\frac{2}{L}|y-z|\right]\labe{hint}\ .\eqne
The $\mu_n$ in these expressions are defined in eq.(\ref{mudef}) and the
matrices ${\cal G}$ are related to the perpendicular
Gauss' law operators in the following way
\dia {\cal G}_{nm}(x) = G_\perp^{c_1}(x)
\frac{\lambda^{c_1}_{nm}}{2}e^{-i(\mu_n-\mu_m)x_3/L}\die
which shows that $H_{int}$ contains singularities only at those points
$x_3^\perp$ where $\mu_n(x_3^\perp)$ and $\mu_m(x_3^\perp)$ coincide for
some $n\neq m$. Using eq.(\ref{jcon}) and the identities given in the appendix
we eventually find in agreement with the results in \cite{Lenz} for the
hamiltonian in the physical Hilbert space the form
\eqna H\!&\!=\!&\! \int d^dx\left\{ \sum_{i\neq 3} \left[\bar\psi(x)V_i(x)
\right]\gamma_i i\partial_i \left[ V_i^{\dag}(x)\psi(x)\right] +\left[
\bar\psi(x) e^{i\theta(x_3^\perp)\frac{x_3}{L}}\right] \gamma_3i\partial_3
\left[ e^{-i\theta(x_3^\perp)\frac{x_3}{L}}\psi(x) \right]
\right. \nonumber \\  && +\frac{g^2}{2} \sum_{i\neq 3} \int_{x_i}^Ldz_i
J_i^b(x_i^\perp,z_i)\int_{x_i}^L dz_i^\prime J_i^b(x_i^\perp,z_i^\prime)
+{\cal K}_3(x_3^\perp) \labe{htrans}\\  && +\left.\fraz \sum_{i,j\neq 3}
\tr{F_{ij}(x)F_{ij}(x)}+ \sum_{i\neq 3} \tr{\tilde F_{i3}(x)\tilde F_{i3}(x)}
+m\bar\psi(x)\psi(x)\right\} + H_{int} \ . \nonumber\eqne
$H_{int}$ has been given in eq.(\ref{hint}) and ${\cal K}_3$
denotes the kinetic energy operator for the fields $\theta^{c_0}(x_3^\perp)$
which is of the form derived in eq.(\ref{kres})
\eqna {\cal K}_3(x_3^\perp) &=& \prod_{k<l}
\sin^{-2}\fraz\left[ \mu_k(x_3^\perp)-\mu_l(x_3^\perp)\right]
\frac{-i\delta}{\delta\theta^{c_0}(x_3^\perp)} \nonumber \\  &&
\cdot\prod_{n<m}
\sin^2\fraz\left[ \mu_n(x_3^\perp)-\mu_m(x_3^\perp)\right]
\frac{-i\delta}{\delta\theta^{c_0}(x_3^\perp)} \ . \eqne
Finally $\tilde F_{i3}$ in the hamiltonian is obtained from
$F_{i3}$ by the replacement
\eqa A_3(x)\rightarrow \frac{i}{g}e^{i\theta(x_3^\perp)x_3/L}\partial_3
e^{-i\theta(x_3^\perp)x_3/L}\ . \eqe
This result shows that all non--trivial properties, including the singular
'Coulomb'--interaction and the Jacobian for the variables $\theta^{c_0}$ are
correctly reproduced. Moreover all operations necessary for this purpose
can be performed on the operator level. Projections onto the physical subspace
have been used in formulating eq.(\ref{htrans}) but exclusively in order to
obtain simple expressions in which irrelevant terms do not show up anymore.
At no stage in the derivation this projection was essential. As discussed
in detail in \cite{Lenz,Ohta} this offers the possibility to transform the
hamiltonian within the physical subspace from one representation to another.
It may also help to clarify the properties of residual gauge transformations
and their relation to the Gribov problem.

Furthermore one can use the formulation in terms of angular variables together
with the technique of unitary gauge fixing transformations to investigate
new gauges. This may allow one to find an optimal starting point for
approximating the full dynamics and thus for extracting physical information
on the low energy properties of QCD. This question is presently under
investigation.\\
\vspace{.5cm}
{\bf Acknowledgments}\\
We would like to thank Profs. K.Ohta, M.Thies and K.Yazaki for helpful
discussions. Support by the Japan Society for the Promotion of Science
is gratefully acknowledged.

\appendix
\section{Useful identities for the unitary transformations}
In this appendix we give useful identities for the derivation of the
hamiltonian in the physical Hilbert space eq.(\ref{htrans}). For the
commutators of $J_3$ with $\cal G$ we find
\eqa \left[J_3^a(x_3^\perp,L),{\cal G}^{c_1}(z)\right] =
\delta^{d-1}(x_3^\perp-z_3^\perp) if^{b_0c_1b_1}R_{ab_0}(x_3^\perp)
{\cal G}^{b_1}(z)\frac{z_3}{L}\eqe
and with the matrices R and P the following commutation relations are obtained
\eqnaa  \left[J_3^a(x_3^\perp,L), R_{bc}(z_3^\perp)\right] &=&
\delta^{d-1}(x_3^\perp-z_3^\perp)if^{c_1cd}R_{bd}(x_3^\perp)
R_{ab_1}(x_3^\perp) P_{b_1c_1}^{-1}(x_3^\perp) \\
\left[J_3^a(x_3^\perp,L),P_{c_1d_1}^{-1}(z_3^\perp)\right] &=&
-\delta^{d-1}(x_3^\perp-z_3^\perp) if^{b_0b_1e_1}R_{ab_0}(x_3^\perp)
P_{b_1d_1}^{-1}(x_3^\perp)\left[\delta_{c_1,e_1}
+P_{c_1,e_1}^{-1}(x_3^\perp)\right]\ . \eqnae
To isolate the physical part of the operator $J_3^aJ_3^a$ we note that the
constraint eq.(\ref{jcon}) suggests that $R_{ab_0}J_3^a$, which is not subject
of this constraint, may be used as a derivative operator with respect to
$\theta^{b_0}$. This is verified by observing the identities
\eqna R_{c_0a}^{-1}(x_3^\perp)\left[J_3^a(x_3^\perp,L),e^{i\theta(z_3^\perp)}
\right] &=& \delta^{d-1}(x_3^\perp-z_3^\perp)e^{i\theta(z_3^\perp)}
\frac{\lambda^{c_0}}{2} \labe{sn} \\
R_{c_0a}^{-1}(x_3^\perp)\left[J_3^a(x_3^\perp,L),e^{i\Delta(z_3^\perp)}
\right] &=& 0\nonumber\\
\Rightarrow R_{c_0a}^{-1}(x_3^\perp)J_3^a(x_3^\perp,L)
&=& \frac{-i\delta}{\delta\theta^{c_0}(x_3^\perp)}\labe{tder}\ .\eqne
Introducing the following notation for $\theta^{c_0}$
using cartesian unit vecors $\vec e_n$ in N dimensions
\eqa \sum_{c_0}\theta^{c_0}(x_3^\perp)\frac{\lambda^{c_0}}{2} =
\sum_{n=1}^N\mu_n(x_3^\perp)\vec e_n\vec e_n^{\dag}\labe{mudef}\eqe
we finally obtain as part of the contribution of the electric field $E_3$ to
the hamiltonian
\eqna J_3^a(x_3^\perp,L)J_3^a(x_3^\perp,L) &=& R_{c_1b}^{-1}(x_3^\perp)J_3^b
(x_3^\perp,L)R_{c_1c}^{-1}(x_3^\perp)J_3^c(x_3^\perp,L) \nonumber \\
&-& \frac{\delta^2}{
\delta\theta^{c_0}(x_3^\perp)\delta\theta^{c_0}(x_3^\perp)}\labe{kres}\\
&-& \frac{\delta^{d-1}(0)}{2}\sum_{n<m}\cot\left[ \frac{\mu_n(x_3^\perp)-
\mu_m(x_3^\perp)}{2} \right] \left(\lambda^{c_0}_{nn}-\lambda^{c_0}_{mm}\right)
\frac{\delta}{\delta\theta^{c_0}(x_3^\perp)}\nonumber \eqne
which gives rise to the operator ${\cal K}_3$ in the hamiltonian acting on
the physical Hilbert space eq.(\ref{htrans}).

\end{document}